\documentclass[twocolumn,aps,prd]{revtex4}
\usepackage{graphicx,amssymb}
\usepackage[utf8]{inputenc} 

\usepackage[urlcolor=blue,colorlinks,breaklinks=true]{hyperref}
\usepackage[anythingbreaks]{breakurl}

\PassOptionsToPackage{obeyspaces,spaces,hyphens}{url}

\usepackage{amsmath}

\begin{document}
\def\half{{1\over2}}
\def\shalf{\textstyle{{1\over2}}}

\newcommand\lsim{\mathrel{\rlap{\lower4pt\hbox{\hskip1pt$\sim$}}
		\raise1pt\hbox{$<$}}}
\newcommand\gsim{\mathrel{\rlap{\lower4pt\hbox{\hskip1pt$\sim$}}
		\raise1pt\hbox{$>$}}}

\title{Second law of thermodynamics in nonminimally coupled gravity}

\author{R.P.L. Azevedo}
\email[Electronic address: ]{rplazevedo@fc.up.pt}
\affiliation{Instituto de Astrof\'{\i}sica e Ci\^encias do Espa{\c c}o, Universidade do Porto, CAUP, Rua das Estrelas, PT4150-762 Porto, Portugal}
\affiliation{Centro de Astrof\'{\i}sica da Universidade do Porto, Rua das Estrelas, PT4150-762 Porto, Portugal}
\affiliation{Departamento de F\'{\i}sica e Astronomia, Faculdade de Ci\^encias, Universidade do Porto, Rua do Campo Alegre 687, PT4169-007 Porto, Portugal}

\author{P.P. Avelino}
\email[Electronic address: ]{pedro.avelino@astro.up.pt}
\affiliation{Instituto de Astrof\'{\i}sica e Ci\^encias do Espa{\c c}o, Universidade do Porto, CAUP, Rua das Estrelas, PT4150-762 Porto, Portugal}
\affiliation{Centro de Astrof\'{\i}sica da Universidade do Porto, Rua das Estrelas, PT4150-762 Porto, Portugal}
\affiliation{Departamento de F\'{\i}sica e Astronomia, Faculdade de Ci\^encias, Universidade do Porto, Rua do Campo Alegre 687, PT4169-007 Porto, Portugal}
\affiliation{School of Physics and Astronomy, University of Birmingham,Birmingham, B15 2TT, United Kingdom}

\date{\today}
\begin{abstract}	
In the present work we show that the second law of thermodynamics does not generally hold if the matter and gravitational fields are nonminimally coupled. We demonstrate this result by explicitly computing the evolution of the entropy of the matter fields in the case of a closed homogeneous and isotropic universe filled with dust and radiation, showing that, in this case, the sign of the entropy variation is determined by the evolution of the universe. The preservation of the second law of thermodynamics in these modified theories would require its generalization to account for a gravitational entropy contribution.
\end{abstract}
\maketitle

\section{Introduction}
\label{sec:intr}

It is well known that in general relativity (GR) the comoving entropy of a homogeneous and isotropic universe filled with a perfect fluid remains constant throughout its evolution, regardless of the fluid's equation of state or the presence (or lack) of curvature. However, comoving entropy conservation may no longer hold in theories featuring a nonminimal coupling (NMC) between geometry and the matter fields \cite{Nojiri2004,Allemandi:2005qs,Bertolami2007,Sotiriou2008}, since in this case the energy momentum tensor is not in general covariantly conserved \cite{Avelino2018,Azevedo2018a,Azevedo2019a}.

In Refs. \cite{Harko2015,Azevedo2019a} it has been shown that the NMC between gravity and the matter fields generally leads to a modification of the first law of thermodynamics. In this context,  gravitationally induced particle creation \cite{Harko2015} has been suggested as a natural consequence of the NMC between matter and geometry. However, more recently it has been shown, using the correct form for the Lagrangian of the matter fields, that particle creation or decay is not a natural consequence of NMC theories, at least in the absence of significant metric perturbations on microscopic scales \cite{Azevedo2019a}. In this paper we will follow up on this work, by showing that the second law of thermodynamics does not generally hold in the context of NMC gravity, and consider a possible link between the evolution of the universe and the arrow of time as defined by the second law of thermodynamics.

The structure of this paper is as follows. In Sec.~\ref{sec:model} we give a brief introduction to NMC gravity and discuss the corresponding changes to the first law of thermodynamics. In Sec.~\ref{sec:entropy} we determine the equation for the evolution of the entropy of the matter fields throughout the evolution of homogeneous and isotropic universes with a NMC between gravity and the matter fields. These results are explicitly computed in Sec.~\ref{sec:background} for the case of a universe with positive curvature filled entirely with dust and radiation. The impact of our results is also discussed in this section, particularly in the light of the second law of thermodynamics. Finally, we conclude in Section~\ref{sec:conc}. We use units such that $c=(16\pi G)^{-1}=1$, where $c$ is the value of the speed of light in vacuum, and $G$ is Newton's gravitational constant. We also adopt the metric signature $(-,+,+,+)$, and the Einstein summation convention.

\section{Nonminimally coupled gravity}
\label{sec:model}

While there are many types of NMC theories featuring complex geometric terms such as  $f(R,{T^{\mu}}_\mu,R_{\mu\nu}T^{\mu\nu})$ gravity \cite{Odintsov2013,Haghani2013}, we consider a simpler model inspired by $f(R)$ theories of gravity due to its explanatory power, and for allowing for the avoidance of  the Ostrogradsky and Dolgov-Kawasaki instabilities \cite{Bertolami2009,Ayuso2015}. It is described by the action
\begin{equation}
\label{eq:action}
S=\int d^4 x \sqrt{-g} \left[ f_1(R) + f_2(R)\mathcal{L}_m\right]\,,
\end{equation}
where $g$ is the determinant of the metric $g_{\mu\nu}$, $\mathcal{L}_m$ is the Lagrangian of the matter fields, and $f_1(R)$ and $f_2(R)$ are generic functions of the Ricci scalar $R$. GR is recovered if $f_1(R)=R$ and $f_2(R)=1$. Assuming a Levi-Civita connection it is straightforward to obtain the equations of motion for the gravitational field
\begin{equation}
\label{eq:eqmotion}
F G_{\mu\nu}=\half f_2 T_{\mu\nu}+\Delta_{\mu\nu}F+\half f_1 g_{\mu\nu}-\half RFg_{\mu\nu}\, ,
\end{equation}
where $G_{\mu\nu}=R_{\mu\nu}-\shalf g_{\mu\nu} R $ is the Einstein tensor, $R_{\mu\nu}$ is the Ricci tensor, $\Delta_{\mu \nu} \equiv \nabla_\mu \nabla_\nu - g_{\mu \nu} \Box$, $\Box \equiv \nabla^\mu \nabla_\mu$,  
\begin{equation}
\label{eq:F}
F= f'_1(R)+f'_2(R)\mathcal{L}_m\,,
\end{equation}
a prime denotes a derivative with respect to the Ricci scalar, and the energy-momentum tensor has the usual form
\begin{equation}
\label{eq:energymom}
T_{\mu\nu}=-{2\over \sqrt{-g}}{\delta(\sqrt{-g}\mathcal{L}_m)\over \delta g^{\mu\nu}}\,.
\end{equation}

Taking the covariant derivative of Eq.~\eqref{eq:eqmotion} and using the Bianchi identities one obtains the following relation in lieu of the usual energy-momentum conservation equation
\begin{equation}
\label{eq:noncons}
\nabla^\mu T_{\mu\nu}={f'_2\over f_2}(g_{\mu\nu}\mathcal{L}_m-T_{\mu\nu})\nabla^\mu R\, .
\end{equation}
Equation~\eqref{eq:noncons} implies that the form of the matter Lagrangian directly affects not only energy-momentum conservation, but also particle motion \cite{Bertolami2008a,Avelino2018}. 

In previous work \cite{Avelino2018a,Avelino2018} it was determined that the on-shell Lagrangian of a perfect fluid composed of noninteracting particles with fixed mass and structure, i.e. solitons, is given by
\begin{equation}
\label{eq:lagrangian}
\mathcal{L}_m={T^{\mu}}_\mu=3p-\rho\,,
\end{equation}
where the energy-momentum tensor is given by
\begin{equation}
\label{eq:pfemt}
T^{\mu\nu}=(\rho+p)u^\mu u^\nu + p g^{\mu\nu}\,,
\end{equation}
$\rho$ and $p$ are the respectively the proper energy density and pressure of a perfect fluid, and $u^\mu$ is its four-velocity. It is noteworthy that the particular structure of the particles is not relevant for this derivation, as long as they do not experience fundamental changes to their structure or mass as a result of the NMC to gravity \cite{Uzan2011,Copeland2007}.

A homogeneous and isotropic universe is described by the flat Friedmann-Lemaître-Robertson-Walker (FLRW) metric with line element
\begin{equation}	
\label{eq:metric}
ds^2=-dt^2+a^2(t)\left[{dr^2 \over1-kr^2} + r^2d\theta^2 +r^2\sin^2\theta d\phi^2\right]\,,
\end{equation} 
where $a(t)$ is the scale factor, $k$ is the curvature, $t$ is the physical time, and $r$, $\theta$, and $\phi$ are spherical comoving spatial coordinates. We take $a(t)$ to be dimensionless, in which case $[k]=[r]^{-2}$ (the brackets represent the dimensions of the corresponding physical quantities). In the remainder of this work we shall use spatial units such that $k=1$.

The dynamics of solitonic particles has been studied in \cite{Sousa2011a,Sousa2011b,Avelino2016}, where it has been shown that in a $3+1$-dimensional FLRW spacetime (ignoring interactions other than gravitational) it is  given by
\begin{equation}
\label{eq:dotvel}
{\dot v} +3\left( H + \frac{{\dot f}_2}{f_2} \right) (1-v^2) v =0 \,.
\end{equation}
where $H\equiv \dot{a}/a$ is the Hubble parameter, $v$ is the speed of the particles, and a dot represents a derivative with respect to the physical time. Hence, the linear momentum of such particles evolves as
\begin{equation}
m \gamma v \propto (a f_2)^{-1}\,, \label{eq:momev}
\end{equation}
where $\gamma \equiv (1-v^2)^{-1/2}$.

The 0th component of Eq.~\eqref{eq:noncons} is given by
\begin{equation}
\label{eq:densityevolution}
{\dot \rho} +  3 H (\rho +p) =  -({\mathcal L}_m  + \rho) \frac{\dot f_2}{f_2} \, .
\end{equation} 
For eqs. \eqref{eq:momev} and \eqref{eq:densityevolution} to be consistent with  the dependence of the proper pressure $p$ on the proper density $\rho$ and on the root mean square velocity of the particles $\langle v^2 \rangle$ ($p=\rho \langle v^2 \rangle /3$), one can easily verify that the matter Lagrangian must indeed given by Eq.~\eqref{eq:lagrangian} (see \cite{Avelino2018a,Avelino2018} for alternative derivations of the same result). Hence, Eq. \eqref{eq:densityevolution} may also be written as
\begin{equation}
\label{eq:densityevolution1}
{\dot \rho} +  3 H (\rho +p) =  -3p \frac{\dot f_2}{f_2} \, .
\end{equation} 

Equation~\eqref{eq:densityevolution} is akin to the first law of thermodynamics,
\begin{equation}
\label{eq:conservenergy}
d (\rho a^3)+pd(a^3)= dQ_{\rm NMC} \, ,
\end{equation}
with the right-hand side acting as a ``heat'' transfer rate per comoving volume, and so we can rewrite it as \cite{Azevedo2019a}
\begin{equation}
\label{eq:heat1}
\dot{\rho}+3H(\rho+p)={\dot{Q}_{\rm NMC}\over a^3} \, .
\end{equation}

While for dust $\dot{Q}_{\rm NMC[dust]} = 0$ (since $p_{\rm dust}=0$), the evolution of the energy density of relativistic matter and radiation is strongly impacted by the NMC coupling to gravity. In particular, this coupling has been shown to lead to a new source of spectral distortion ($n$-type spectral distortions) of the CMB power spectrum Ref.~\cite{Avelino2018}. The energy-momentum nonconservation associated to the NMC to gravity suggests that a more general definition of the energy-momentum tensor, including a yet to be defined gravitational contribution, should be considered. This has not yet been sufficiently explored in the context of NMC theories, but has proven to be quite problematic in the context of GR \cite{Bonilla1997,Clifton2013,Sussman2014}.

\section{The second law of thermodynamics}\label{sec:entropy}

Consider the fundamental thermodynamic relation
\begin{equation}
\label{eq:ftr}
TdS = d(\rho a^3) + pda^3\, ,
\end{equation}
where $S$ is the entropy of the matter content and $T$ is the temperature. Equations \eqref{eq:densityevolution1}, \eqref{eq:heat1} and \eqref{eq:ftr} imply that
\begin{equation}
\label{eq:entropy}
TdS = dQ_{\rm NMC}= -3p a^3 \frac{d f_2}{f_2} \, ,
\end{equation}
as opposed to GR, where $dS = 0$.

Equation {\eqref{eq:entropy} implies that the comoving entropy of the dust component is conserved ($dS_{\rm dust}=0$ since $p_{\rm dust}=0$). In this case the total comoving entropy $S$ is equal to the comoving entropy of the radiation component $S_{\rm r}$ ($S=S_{\rm r}$). Its proper pressure and density --- be it bosonic, fermionic or both ---- satisfy $p_{\rm r}=\rho_{\rm r}/3$ and $\rho_{\rm r} \propto T^4$, respectively --- here, the scattering timescale is implicitly assumed to be much smaller than the characteristic timescale of the change of the NMC coupling function $f_2$, so that the radiation component can always be taken to be in thermodynamic equilibrium at a temperature $T$ (we are also assuming that the chemical potential is zero). Hence, in the case of radiation, Eq.~\eqref{eq:densityevolution1} may be written as
\begin{equation}
\label{eq:conservenergypart}
\frac{d\rho_{\rm r}}{dT}\, \dot{T}+3H(\rho_{\rm r}+p_{\rm r})=-3p_{\rm r}\frac{\dot{f}_2}{f_2} \,,
\end{equation}
or equivalently,
\begin{equation}
\label{eq:Tdot1}
\dot{T}=- \frac{3H\left(\rho_{\rm r}+p_{\rm r}\right)+3p_{\rm r}\frac{\dot{f}_2}{f_2}}{{\frac{d\rho_{\rm r}}{dT}}} \,,
\end{equation}
with $p_{\rm r}=\rho_{\rm r}/3$ and $\rho_{\rm r} \propto T^4$.

Equation~\eqref{eq:Tdot1} is easily integrated and returns
\begin{equation}
\label{eq:Tradeq}
T \propto a^{-1} f_2^{-1/4} \,,
\end{equation}
so that 
\begin{equation}
\label{eq:densrad}
\rho_{\rm r} \propto a^{-4}f_2^{-1} \,.
\end{equation}
Taking into account Eq. \eqref{eq:Tradeq} and the fact that $p_{\rm r}=\rho_{\rm r}/3 \propto T^4$, Eq.~\eqref{eq:entropy} can be easily integrated to give
\begin{equation}
\label{eq:entropyevoeq}
S\propto f_2^{-3/4} \,.
\end{equation}

Imposing the second law of thermodynamics
\begin{equation}
\label{eq:entropycreation}
T\dot{S} = \dot{Q}_{\rm NMC} = -3p a^3\frac{f'_2}{f_2}\dot{R} \geq 0\, ,
\end{equation}
would prove quite restrictive, in particular in the case of a universe in which the time derivative of the Ricci scalar changes sign, as we will demonstrate in the next section. In fact, the only function $f_2$ that would verify Eq. \eqref{eq:entropycreation} with all generality would be $f_2=\text{const.}$, which corresponds to the GR limit.

\section{Numerical Results and Discussion}\label{sec:background}

Here, we shall consider a homogeneous and isotropic universe with positive curvature filled with dust and radiation. This provides a fair representation of the energy content of the Universe from early post-inflationary times until the onset of dark energy domination. The addition of a positive curvature will allow us to consider expanding and contracting phases of evolution of the universe and to contrast the behavior of the comoving entropy of the matter fields in these periods.

The total proper energy density and pressure are given, respectively, by
\begin{equation}
\label{eq:densitypressure}
\rho_{\rm total}=\rho_{\rm r}+\rho_{\rm dust}\,, \qquad \qquad p_{\rm total} =\frac{1}{3}\rho_{\rm r} \,,
\end{equation}
with
\begin{equation}
\label{eq:realdensities}
\rho_{\rm r} =\rho_{\rm r0} \, a^{-4}f_2^{-1}\,, \qquad \qquad \rho_{\rm dust} =\rho_{\rm dust} \, a^{-3} \,.
\end{equation}
On the other hand, Eqs. \eqref{eq:lagrangian} , \eqref{eq:densitypressure}, and \eqref{eq:realdensities},  imply that the Lagrangian of the matter fields is equal to
\begin{equation}
\label{eq:lagrangianrpcdm}
\mathcal{L}_m=3p_{\rm total}-\rho_{\rm total}=-\rho_{\rm dust0} \, a^{-3}\,.
\end{equation}
Here, the subscripts `$\rm r$' and `$\rm dust$' again denote the radiation and cold dark matter components, respectively, and the subscript `$0$' refers to an arbitrary initial time $t=0$. 

By evaluating the $tt$  and $ii$ components of Eq.~\eqref{eq:eqmotion}, where `$i$' represents any of the spatial coordinates, and using Eqs. \eqref{eq:F}, \eqref{eq:densitypressure}, \eqref{eq:realdensities}, \eqref{eq:lagrangianrpcdm}, along with
\begin{align}
\label{eq:deltaF}
& \Delta_{tt} F = -3HF' \dot{R} - 9H^2f'_2 \rho_{\rm dust} \nonumber\\
&=-18HF' \left(\ddot{H}+4H\dot{H} -  2Ha^{-2} \right) - 9H^2f'_2 \rho_{\rm dust}\,,\\
&\Delta_{ii}F=g_{ii}\left(2H\dot{F}+\ddot{F}\right)\,,
\end{align}
where the time derivatives of $F$ are
\begin{align}
\label{eq:Fdot}
\dot{F}&=F'\dot{R} + 3H f'_2 \rho_{\rm dust}\,, \\
\ddot{F} &= F'\ddot{R}+F''\dot{R}^2 +3\left(2H\dot{R}f''_2+\dot{H}f'_2-3H^2f'_2\right)\rho_{\rm dust}\,,
\end{align}
and
\begin{align}
\label{eq:R}
R&=6\left(\dot{H}+2H^2+a^{-2}\right)\,, \\
\dot{R}&=6\left(\ddot{H}+4H\dot{H}-2Ha^{-2}\right)\,, \\
\ddot{R}&=6\left[\dddot{H}+4H\ddot{H}+4\dot{H}^2+2\left(2H^2-\dot{H}\right)a^{-2}\right] \,,
\end{align}
it is straightforward to obtain the modified Friedmann and Raychaudhuri equations
\begin{align}
\label{eq:friedmann}
\left(H^2 +a^{-2}\right)F =& \frac{1}{6}\left(f'_1 R-f_1\right)+ \frac{1}{6} \rho_{\rm r0}a^{-4}- HF'\dot{R}\nonumber\\
&+\frac{1}{6}\left[f_2-f'_2 \left(R+18H^2\right)\right]  \rho_{\rm dust0}a^{-3} \,,
\end{align}
\begin{align}
\label{eq:ray}
\left(\dot{H}+H^2\right)F=&-2\left(H^2+a^{-2}\right)F+\frac{1}{2} f_1 \nonumber \\
&+\frac{1}{6}\rho_{\rm r0}a^{-4}f_2^{-1} +2H\dot{F} +\ddot{F} \,,
\end{align}
which are respectively third and fourth order nonlinear differential equation for the scale factor $a$ with respect to time (due to the $\dot{R}$ and $\ddot{F}$ terms).

Here, we consider the functions $f_1 = R$ and $f_2 = \alpha R^\beta$, with constant $\alpha$ and $\beta$, so that Eq. (\ref{eq:friedmann}) becomes
\begin{align}
\label{eq:friedmann1}
&\left(H^2 +a^{-2}\right)F = \frac{1}{6} \rho_{\rm r0}a^{-4}\nonumber\\
&+ \frac{\alpha}{6} R^{\beta} \left(1-\beta- \frac{18H^2}{R}\right)  \rho_{\rm dust0}a^{-3} \nonumber\\
& + 6 \alpha \beta (\beta-1) R^{\beta-2}  \rho_{\rm dust0}a^{-3} H^2 \left(\frac{\ddot{H}}{H}+4\dot{H}-2a^{-2}\right)\,.
\end{align}
Starting from an arbitrary initial time ($t=0$), we integrate Eq. \eqref{eq:ray} using a 5th-order backwards differentiation formula, first backwards up to the Big-Bang and then forward up to the Big Crunch. Since it is a fourth order differential equation it requires setting three further initial conditions ($H_0$, $\dot{H}_0$ and $\ddot{H}_0$) in addition to $a_0=1$ (as well as $\rho_{\rm dust0}$ and $\rho_{\rm r0}$).

Notice that the comoving entropy may change only if $\rho_{\rm dust} \neq 0$ and $\rho_{\rm r} \neq 0$. If the universe was assumed to be filled entirely with cold dark matter, then the  proper pressure would vanish and, therefore, so would the right-hand side of Eq.~\eqref{eq:entropy}. Hence, there would be no change to the comoving entropy content of the universe. Conversely, if the universe was composed only of radiation, then  Eq.~\eqref{eq:friedmann1} would reduce to the standard Friedmann equation found in GR. Hence, the Ricci scalar $R$ would vanish and, Eq.~\eqref{eq:entropy} would again imply the conservation of the comoving entropy. 

\begin{figure}
	\centering
	\includegraphics[width=\columnwidth]{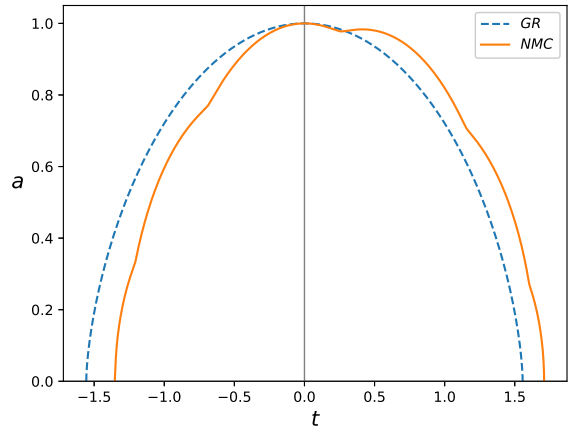}
	\caption{Evolution of the scale factor $a$ as a function of the physical time $t$ in the context of GR (blue dashed line) and of a NMC gravity model with $\alpha=0.95$ and $\beta=0.01$ (orange solid line), having $\rho_{\rm dust0}=5.94$ and $\rho_{\rm r0}=0.06$ and $H_0=0$ as initial conditions. An extra initial condition is required in the context of NMC gravity, which we take to be ${\ddot H}_0=0.5$. Notice the asymmetric evolution of the universe in the context of NMC gravity (in contrast with GR), and the presence of oscillations of variable amplitude and frequency, as well as two local maxima of $a$.
\label{fig:asyma}}
\end{figure}

In the remainder of this paper we shall consider cosmologies with $\rho_{\rm dust0}=5.94$, $\rho_{\rm r0}=0.06$ and $H_0=0$ in the context either of GR or of NMC gravity models with  $\alpha=0.95$ and $\beta=0.01$. In the case of GR ($\alpha=1$, $\beta=0$), these conditions are sufficient to determine the full evolution of the universe. In the context of NMC gravity, Eq.~\eqref{eq:friedmann1} acts as an additional constraint, and with $a_0=1$ and $H_0=0$ becomes
\begin{equation}
\label{eq:MFEcond}
\rho_{\rm r0}+\alpha\left[R_0^\beta+\beta(6-R_0)R_0^{\beta-1} \right]\rho_{\rm dust0}
=6 \,,
\end{equation}
and therefore sets $\dot{H}_0$ at the initial time, leaving only one additional initial condition, $\ddot{H}_0$ .

Fig.~\ref{fig:asyma} displays the evolution of the scale factor $a$ as a function of the physical time $t$ in the context of two distinct cosmological models computed assuming either GR (blue dashed line) or  NMC gravity (orange solid line). In the context of GR one may observe the exact symmetry between the expanding and contracting phases of the universe, which is verified independently of the initial conditions. The orange solid line shows the evolution of $a$ with $t$ in the context of a NMC gravity model with  $\ddot{H}_0=0.5$. Fig.~\ref{fig:asyma} shows that, in this case, the symmetry between the expanding and contracting phases of the universe is no longer preserved. It also reveals the presence of oscillations on the evolution of the scale factor of variable amplitude and frequency, as well as multiple (two, for this particular parameter choice) local maxima of the scale factor. These features are common in the context of NMC gravity, and are associated with the increased complexity of the higher order nonlinear equations which rule the evolution of the universe in that context. Moreover, many NMC models (such as the present one for the chosen parameters) are subject to the Dolgov-Kawasaki instability \cite{Faraoni2007,Bertolami2009a}. However, that oscillatory behavior in the cosmological evolution of the universe has also been previously discussed for $f(R)$ models \cite{Appleby2010,Motohashi2011}, even when they satisfy the former and other stability criteria. A detailed analysis of such oscillations is not the main purpose of this paper, as they do not affect its main result --- that the second law of thermodynamics does not generally hold in the context of NMC gravity. Fig.~\ref{fig:asymH} displays the evolution of the Hubble parameter $H$ as a function of the physical time $t$ for the same models shown in Fig.~\ref{fig:asyma}. Notice the three zeros of $H$, as well as its sharp variations at specific values of the physical time $t$ in the case of NMC gravity.

\begin{figure}
	\centering
	\includegraphics[width=\columnwidth]{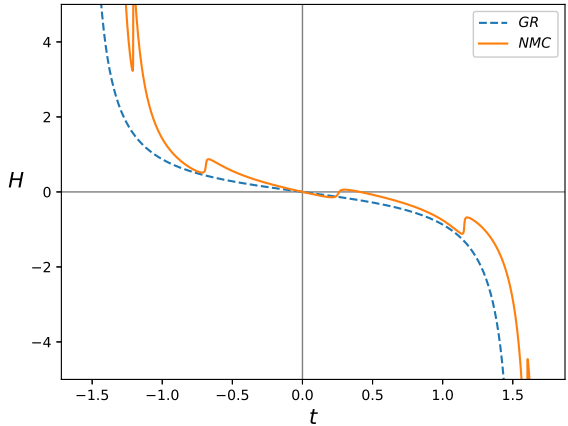}
	\caption{Same as in Fig.~\ref{fig:asyma} but for the evolution of the Hubble parameter $H$. Notice the three zeros of $H$, as well as its sharp variation at specific values of the physical time $t$ in the context of NMC gravity. \label{fig:asymH}}
\end{figure}

Although an asymmetry between the expanding and contracting phases is generic in the context of NMC gravity, one  can use the freedom in the choice of initials conditions to impose a symmetric expansion and contraction by choosing $\ddot{H}_0 = 0$.

\begin{figure}
	\centering
	\includegraphics[width=\columnwidth]{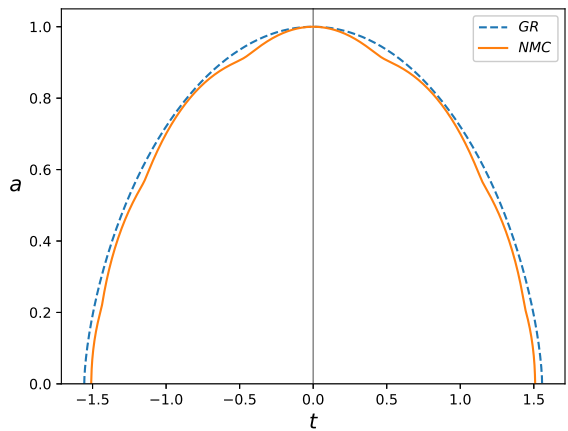}
	\caption{Evolution of the scale factor $a$ as a function of the physical time $t$ in the context of GR (blue dashed line) and of a NMC gravity model with $\alpha=0.95$ and $\beta=0.01$ (orange solid line), having $\rho_{\rm dust0}=5.94$, $\rho_{\rm r0}=0.06$ and $H_0=0$ as initial conditions. An extra initial condition is required in the context of NMC gravity which we take to be ${\ddot H}_0=0$ in order to guarantee a symmetric evolution of the universe. 
\label{fig:syma}}
\end{figure}
\begin{figure}
	\centering
	\includegraphics[width=\columnwidth]{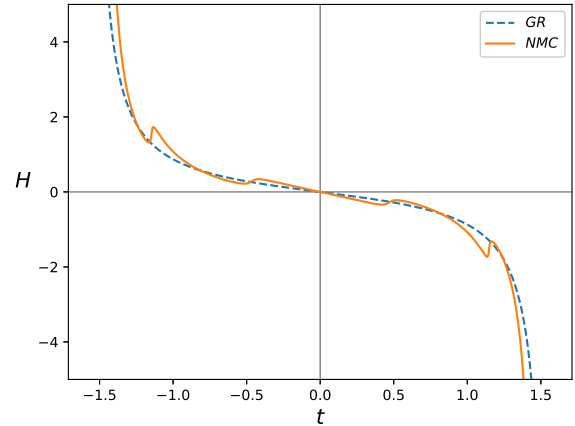}
	\caption{Same as in Fig.~\ref{fig:syma} but for the evolution of the Hubble parameter $H$. \label{fig:symH}}
\end{figure}
\begin{figure}
	\centering
	\includegraphics[width=\columnwidth]{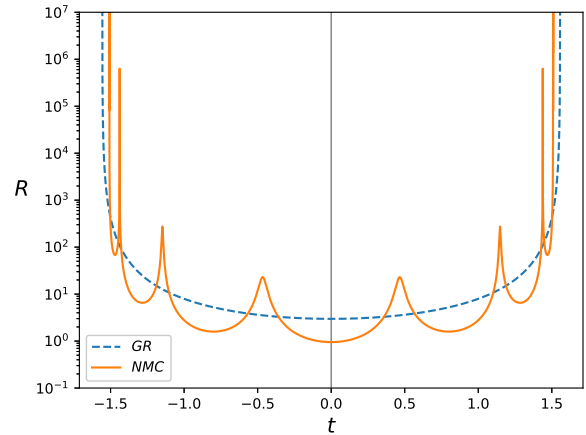}
	\caption{Same as in Fig.~\ref{fig:syma} but for the evolution of the Ricci scalar $R$. \label{fig:symR}}
\end{figure}
\begin{figure}
	\centering
	\includegraphics[width=\columnwidth]{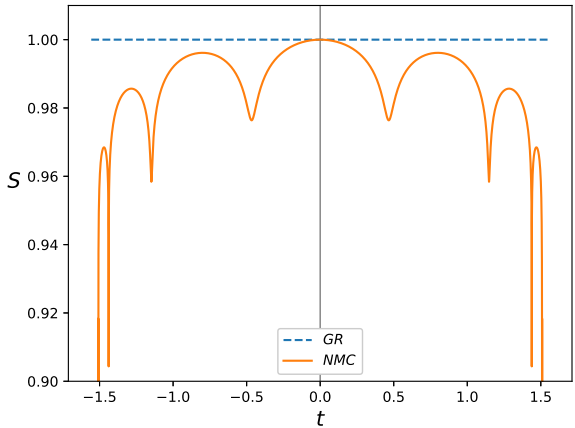}
	\caption{Same as in Fig.~\ref{fig:syma} but for the evolution of the comoving entropy $S$, normalized so that $S=1$ at $t=0$. \label{fig:symS}}
\end{figure}

The results for the symmetric case can be found in Figs.~\ref{fig:syma} through \ref{fig:symS}, which show, respectively, the evolution of the scale factor $a$, the Hubble parameter $H$, the Ricci scalar $R$ and the entropy $S$ as a function of the the physical time $t$. The results presented in Figs.~\ref{fig:syma} and \ref{fig:symH} for the evolution of $a$ and $H$ with the physical time, display an exact symmetry between the expanding and contracting phases of the universe, both in the case of GR and NMC gravity. Also note that in this case there is a single maximum of $a$ (zero of $H$). Otherwise, the results are similar to those shown in Figs.~\ref{fig:asyma} and \ref{fig:asymH} for an asymmetric evolution of the universe. 

Figs.~\ref{fig:symR} and \ref{fig:symS} display the evolution of the Ricci scalar  $R$ and of the comoving entropy $S$ again for the symmetric case. Apart from the oscillations of variable amplitude and frequency, Fig.~\ref{fig:symR} shows that, on average, $R$ decreases during the expanding phase and increases during the contracting phase, while Fig. \ref{fig:symS} show that the comoving entropy has the opposite behavior. This illustrates the coupling between the evolution of the comoving entropy and the dynamics of the universe, which generally exists in cosmological models with a NMC between gravity and the matter fields, linking the thermodynamic arrow of time to the cosmological evolution of the universe.

\section{Conclusions}\label{sec:conc}

In this paper we have shown that the second law of thermodynamics does not generally hold in the presence of a NMC between matter and geometry. We have shown that in a homogeneous and isotropic universe the sign of the entropy variation in NMC gravity is coupled to the evolution of the Ricci scalar and, therefore, to the dynamics of the universe as whole.  Although the evolution of the entropy was computed explicitly only for the case of a universe filled with cold dark matter and radiation, our results are quite generic at least in the absence of significant perturbations of the matter fields.

The preservation of the second law of thermodynamics would require its generalization in order to take into account a gravitational entropy contribution. Even though some work has been done in this context for GR \cite{Bonilla1997,Clifton2013,Sussman2014,Acquaviva2018}, it remains a subject of much discussion and debate, and no clear candidate for a gravitational energy-momentum tensor and gravitational entropy has emerged. However, the need for such a generalized description appears to be much greater in modified gravity models that inherently violate the second law of thermodynamics in its standard form.

\begin{acknowledgments}
	
R.P.L.A. was supported by the Funda{\c c}\~ao para a Ci\^encia e Tecnologia (FCT, Portugal) grant SFRH/BD/132546/2017. P.P.A. acknowledges the support from Fundação para a Ciência e a Tecnologia (FCT) through the Sabbatical Grant No. SFRH/BSAB/150322/2019. Funding of this work has also been provided by FCT through national funds (PTDC/FIS-PAR/31938/2017) and by FEDER—Fundo Europeu de Desenvolvimento Regional through COMPETE2020 - Programa Operacional Competitividade e Internacionaliza{\c c}\~ao (POCI-01-0145-FEDER-031938), and through the research grants UID/FIS/04434/2019, UIDB/04434/2020 and UIDP/04434/2020. This paper benefited from the participation of the authors on the COST action CA15117 (CANTATA), supported by COST (European Cooperation in Science and Technology).
\end{acknowledgments}

\bibliography{NMC_Curvature}

\begin{thebibliography}{27}
\expandafter\ifx\csname natexlab\endcsname\relax\def\natexlab#1{#1}\fi
\expandafter\ifx\csname bibnamefont\endcsname\relax
  \def\bibnamefont#1{#1}\fi
\expandafter\ifx\csname bibfnamefont\endcsname\relax
  \def\bibfnamefont#1{#1}\fi
\expandafter\ifx\csname citenamefont\endcsname\relax
  \def\citenamefont#1{#1}\fi
\expandafter\ifx\csname url\endcsname\relax
  \def\url#1{\texttt{#1}}\fi
\expandafter\ifx\csname urlprefix\endcsname\relax\def\urlprefix{URL }\fi
\providecommand{\bibinfo}[2]{#2}
\providecommand{\eprint}[2][]{\url{#2}}

\bibitem[{\citenamefont{Nojiri and Odintsov}(2004)}]{Nojiri2004}
\bibinfo{author}{\bibfnamefont{S.}~\bibnamefont{Nojiri}} \bibnamefont{and}
  \bibinfo{author}{\bibfnamefont{S.~D.} \bibnamefont{Odintsov}},
  \bibinfo{journal}{Phys. Lett. B} \textbf{\bibinfo{volume}{599}},
  \bibinfo{pages}{137} (\bibinfo{year}{2004}), ISSN \bibinfo{issn}{03702693},
  \eprint{0403622},
  \urlprefix\url{http://linkinghub.elsevier.com/retrieve/pii/S0370269304012201}.

\bibitem[{\citenamefont{Allemandi et~al.}(2005)\citenamefont{Allemandi,
  Borowiec, Francaviglia, and Odintsov}}]{Allemandi:2005qs}
\bibinfo{author}{\bibfnamefont{G.}~\bibnamefont{Allemandi}},
  \bibinfo{author}{\bibfnamefont{A.}~\bibnamefont{Borowiec}},
  \bibinfo{author}{\bibfnamefont{M.}~\bibnamefont{Francaviglia}},
  \bibnamefont{and} \bibinfo{author}{\bibfnamefont{S.~D.}
  \bibnamefont{Odintsov}}, \bibinfo{journal}{Phys. Rev. D}
  \textbf{\bibinfo{volume}{72}}, \bibinfo{pages}{063505}
  (\bibinfo{year}{2005}), ISSN \bibinfo{issn}{1550-7998},
  \urlprefix\url{https://link.aps.org/doi/10.1103/PhysRevD.72.063505}.

\bibitem[{\citenamefont{Bertolami et~al.}(2007)\citenamefont{Bertolami,
  B{\"{o}}hmer, Harko, and Lobo}}]{Bertolami2007}
\bibinfo{author}{\bibfnamefont{O.}~\bibnamefont{Bertolami}},
  \bibinfo{author}{\bibfnamefont{C.~G.} \bibnamefont{B{\"{o}}hmer}},
  \bibinfo{author}{\bibfnamefont{T.}~\bibnamefont{Harko}}, \bibnamefont{and}
  \bibinfo{author}{\bibfnamefont{F.~S.~N.} \bibnamefont{Lobo}},
  \bibinfo{journal}{Phys. Rev. D} \textbf{\bibinfo{volume}{75}},
  \bibinfo{pages}{104016} (\bibinfo{year}{2007}), ISSN
  \bibinfo{issn}{1550-7998}, \eprint{0704.1733},
  \urlprefix\url{https://link.aps.org/doi/10.1103/PhysRevD.75.104016}.

\bibitem[{\citenamefont{Sotiriou and Faraoni}(2008)}]{Sotiriou2008}
\bibinfo{author}{\bibfnamefont{T.~P.} \bibnamefont{Sotiriou}} \bibnamefont{and}
  \bibinfo{author}{\bibfnamefont{V.}~\bibnamefont{Faraoni}},
  \bibinfo{journal}{Class. Quantum Gravity} \textbf{\bibinfo{volume}{25}},
  \bibinfo{pages}{205002} (\bibinfo{year}{2008}), ISSN
  \bibinfo{issn}{0264-9381}, \eprint{arXiv:0805.1249v2},
  \urlprefix\url{http://stacks.iop.org/0264-9381/25/i=20/a=205002?key=crossref.1418b59d85760534de845481e6f1fd48}.

\bibitem[{\citenamefont{Avelino and Azevedo}(2018)}]{Avelino2018}
\bibinfo{author}{\bibfnamefont{P.~P.} \bibnamefont{Avelino}} \bibnamefont{and}
  \bibinfo{author}{\bibfnamefont{R.~P.~L.} \bibnamefont{Azevedo}},
  \bibinfo{journal}{Phys. Rev. D} \textbf{\bibinfo{volume}{97}},
  \bibinfo{pages}{64018} (\bibinfo{year}{2018}), ISSN
  \bibinfo{issn}{2470-0010},
  \urlprefix\url{https://link.aps.org/doi/10.1103/PhysRevD.97.064018}.

\bibitem[{\citenamefont{Azevedo and Avelino}(2018)}]{Azevedo2018a}
\bibinfo{author}{\bibfnamefont{R.~P.~L.} \bibnamefont{Azevedo}}
  \bibnamefont{and} \bibinfo{author}{\bibfnamefont{P.~P.}
  \bibnamefont{Avelino}}, \bibinfo{journal}{Phys. Rev. D}
  \textbf{\bibinfo{volume}{98}}, \bibinfo{pages}{064045}
  (\bibinfo{year}{2018}), ISSN \bibinfo{issn}{2470-0010},
  \urlprefix\url{https://link.aps.org/doi/10.1103/PhysRevD.98.064045}.

\bibitem[{\citenamefont{Azevedo and Avelino}(2019)}]{Azevedo2019a}
\bibinfo{author}{\bibfnamefont{R.~P.~L.} \bibnamefont{Azevedo}}
  \bibnamefont{and} \bibinfo{author}{\bibfnamefont{P.~P.}
  \bibnamefont{Avelino}}, \bibinfo{journal}{Phys. Rev. D}
  \textbf{\bibinfo{volume}{99}}, \bibinfo{pages}{064027}
  (\bibinfo{year}{2019}), ISSN \bibinfo{issn}{2470-0010},
  \urlprefix\url{https://link.aps.org/doi/10.1103/PhysRevD.99.064027}.

\bibitem[{\citenamefont{Harko et~al.}(2015)\citenamefont{Harko, Lobo, Mimoso,
  and Pav{\'{o}}n}}]{Harko2015}
\bibinfo{author}{\bibfnamefont{T.}~\bibnamefont{Harko}},
  \bibinfo{author}{\bibfnamefont{F.~S.~N.} \bibnamefont{Lobo}},
  \bibinfo{author}{\bibfnamefont{J.~P.} \bibnamefont{Mimoso}},
  \bibnamefont{and}
  \bibinfo{author}{\bibfnamefont{D.}~\bibnamefont{Pav{\'{o}}n}},
  \bibinfo{journal}{Eur. Phys. J. C} \textbf{\bibinfo{volume}{75}},
  \bibinfo{pages}{386} (\bibinfo{year}{2015}), ISSN \bibinfo{issn}{1434-6044},
  \eprint{1508.02511},
  \urlprefix\url{http://link.springer.com/10.1140/epjc/s10052-015-3620-5}.

\bibitem[{\citenamefont{Odintsov and
  S{\'{a}}ez-G{\'{o}}mez}(2013)}]{Odintsov2013}
\bibinfo{author}{\bibfnamefont{S.~D.} \bibnamefont{Odintsov}} \bibnamefont{and}
  \bibinfo{author}{\bibfnamefont{D.}~\bibnamefont{S{\'{a}}ez-G{\'{o}}mez}},
  \bibinfo{journal}{Phys. Lett. B} \textbf{\bibinfo{volume}{725}},
  \bibinfo{pages}{437} (\bibinfo{year}{2013}), ISSN \bibinfo{issn}{03702693},
  \eprint{arXiv:1304.5411},
  \urlprefix\url{http://linkinghub.elsevier.com/retrieve/pii/S0370269313005881}.

\bibitem[{\citenamefont{Haghani et~al.}(2013)\citenamefont{Haghani, Harko,
  Lobo, Sepangi, and Shahidi}}]{Haghani2013}
\bibinfo{author}{\bibfnamefont{Z.}~\bibnamefont{Haghani}},
  \bibinfo{author}{\bibfnamefont{T.}~\bibnamefont{Harko}},
  \bibinfo{author}{\bibfnamefont{F.~S.~N.} \bibnamefont{Lobo}},
  \bibinfo{author}{\bibfnamefont{H.~R.} \bibnamefont{Sepangi}},
  \bibnamefont{and} \bibinfo{author}{\bibfnamefont{S.}~\bibnamefont{Shahidi}},
  \bibinfo{journal}{Phys. Rev. D} \textbf{\bibinfo{volume}{88}},
  \bibinfo{pages}{044023} (\bibinfo{year}{2013}), ISSN
  \bibinfo{issn}{1550-7998},
  \urlprefix\url{https://link.aps.org/doi/10.1103/PhysRevD.88.044023}.

\bibitem[{\citenamefont{Bertolami and
  Sequeira}(2009{\natexlab{a}})}]{Bertolami2009}
\bibinfo{author}{\bibfnamefont{O.}~\bibnamefont{Bertolami}} \bibnamefont{and}
  \bibinfo{author}{\bibfnamefont{M.~C.} \bibnamefont{Sequeira}},
  \bibinfo{journal}{Phys. Rev. D} \textbf{\bibinfo{volume}{79}},
  \bibinfo{pages}{104010} (\bibinfo{year}{2009}{\natexlab{a}}), ISSN
  \bibinfo{issn}{1550-7998}, \eprint{0903.4540},
  \urlprefix\url{https://link.aps.org/doi/10.1103/PhysRevD.79.104010}.

\bibitem[{\citenamefont{Ayuso et~al.}(2015)\citenamefont{Ayuso, {Beltr{\'{a}}n
  Jim{\'{e}}nez}, and de~la Cruz-Dombriz}}]{Ayuso2015}
\bibinfo{author}{\bibfnamefont{I.}~\bibnamefont{Ayuso}},
  \bibinfo{author}{\bibfnamefont{J.}~\bibnamefont{{Beltr{\'{a}}n
  Jim{\'{e}}nez}}}, \bibnamefont{and}
  \bibinfo{author}{\bibfnamefont{{\'{A}}.}~\bibnamefont{de~la Cruz-Dombriz}},
  \bibinfo{journal}{Phys. Rev. D} \textbf{\bibinfo{volume}{91}},
  \bibinfo{pages}{104003} (\bibinfo{year}{2015}), ISSN
  \bibinfo{issn}{1550-7998},
  \urlprefix\url{https://link.aps.org/doi/10.1103/PhysRevD.91.104003}.

\bibitem[{\citenamefont{Bertolami et~al.}(2008)\citenamefont{Bertolami, Lobo,
  and P{\'{a}}ramos}}]{Bertolami2008a}
\bibinfo{author}{\bibfnamefont{O.}~\bibnamefont{Bertolami}},
  \bibinfo{author}{\bibfnamefont{F.~S.~N.} \bibnamefont{Lobo}},
  \bibnamefont{and}
  \bibinfo{author}{\bibfnamefont{J.}~\bibnamefont{P{\'{a}}ramos}},
  \bibinfo{journal}{Phys. Rev. D} \textbf{\bibinfo{volume}{78}},
  \bibinfo{pages}{064036} (\bibinfo{year}{2008}), ISSN
  \bibinfo{issn}{1550-7998}, \eprint{0806.4434},
  \urlprefix\url{https://link.aps.org/doi/10.1103/PhysRevD.78.064036}.

\bibitem[{\citenamefont{Avelino and Sousa}(2018)}]{Avelino2018a}
\bibinfo{author}{\bibfnamefont{P.~P.} \bibnamefont{Avelino}} \bibnamefont{and}
  \bibinfo{author}{\bibfnamefont{L.}~\bibnamefont{Sousa}},
  \bibinfo{journal}{Phys. Rev. D} \textbf{\bibinfo{volume}{97}},
  \bibinfo{pages}{64019} (\bibinfo{year}{2018}), ISSN
  \bibinfo{issn}{2470-0010}, \eprint{1802.03961},
  \urlprefix\url{https://link.aps.org/doi/10.1103/PhysRevD.97.064019}.

\bibitem[{\citenamefont{Uzan}(2011)}]{Uzan2011}
\bibinfo{author}{\bibfnamefont{J.-P.} \bibnamefont{Uzan}},
  \bibinfo{journal}{Living Rev. Relativ.} \textbf{\bibinfo{volume}{14}},
  \bibinfo{pages}{2} (\bibinfo{year}{2011}), ISSN \bibinfo{issn}{2367-3613},
  \eprint{1009.5514},
  \urlprefix\url{http://link.springer.com/10.12942/lrr-2011-2}.

\bibitem[{\citenamefont{Copeland et~al.}(2007)\citenamefont{Copeland, Rajantie,
  Contaldi, Dauncey, and Stoica}}]{Copeland2007}
\bibinfo{author}{\bibfnamefont{E.~J.} \bibnamefont{Copeland}},
  \bibinfo{author}{\bibfnamefont{A.}~\bibnamefont{Rajantie}},
  \bibinfo{author}{\bibfnamefont{C.}~\bibnamefont{Contaldi}},
  \bibinfo{author}{\bibfnamefont{P.}~\bibnamefont{Dauncey}}, \bibnamefont{and}
  \bibinfo{author}{\bibfnamefont{H.}~\bibnamefont{Stoica}}, in
  \emph{\bibinfo{booktitle}{AIP Conf. Proc.}} (\bibinfo{publisher}{AIP},
  \bibinfo{year}{2007}), vol. \bibinfo{volume}{957}, pp.
  \bibinfo{pages}{21--29}, ISBN \bibinfo{isbn}{9780735404717}, ISSN
  \bibinfo{issn}{0094243X}, \eprint{0603057},
  \urlprefix\url{http://aip.scitation.org/doi/abs/10.1063/1.2823765}.

\bibitem[{\citenamefont{Sousa and Avelino}(2011{\natexlab{a}})}]{Sousa2011a}
\bibinfo{author}{\bibfnamefont{L.}~\bibnamefont{Sousa}} \bibnamefont{and}
  \bibinfo{author}{\bibfnamefont{P.~P.} \bibnamefont{Avelino}},
  \bibinfo{journal}{Phys. Rev. D} \textbf{\bibinfo{volume}{83}},
  \bibinfo{pages}{103507} (\bibinfo{year}{2011}{\natexlab{a}}), ISSN
  \bibinfo{issn}{1550-7998}, \eprint{arXiv:1103.1381v1},
  \urlprefix\url{https://link.aps.org/doi/10.1103/PhysRevD.83.103507}.

\bibitem[{\citenamefont{Sousa and Avelino}(2011{\natexlab{b}})}]{Sousa2011b}
\bibinfo{author}{\bibfnamefont{L.}~\bibnamefont{Sousa}} \bibnamefont{and}
  \bibinfo{author}{\bibfnamefont{P.~P.} \bibnamefont{Avelino}},
  \bibinfo{journal}{Phys. Rev. D} \textbf{\bibinfo{volume}{84}},
  \bibinfo{pages}{063502} (\bibinfo{year}{2011}{\natexlab{b}}), ISSN
  \bibinfo{issn}{1550-7998}, \eprint{arXiv:1107.4582v1},
  \urlprefix\url{https://link.aps.org/doi/10.1103/PhysRevD.84.063502}.

\bibitem[{\citenamefont{Avelino and Sousa}(2016)}]{Avelino2016}
\bibinfo{author}{\bibfnamefont{P.~P.} \bibnamefont{Avelino}} \bibnamefont{and}
  \bibinfo{author}{\bibfnamefont{L.}~\bibnamefont{Sousa}},
  \bibinfo{journal}{Phys. Rev. D} \textbf{\bibinfo{volume}{93}},
  \bibinfo{pages}{023519} (\bibinfo{year}{2016}), ISSN
  \bibinfo{issn}{2470-0010}, \eprint{1511.05897},
  \urlprefix\url{https://link.aps.org/doi/10.1103/PhysRevD.93.023519}.

\bibitem[{\citenamefont{Bonilla and Senovilla}(1997)}]{Bonilla1997}
\bibinfo{author}{\bibfnamefont{M.~{\'{A}}.~G.} \bibnamefont{Bonilla}}
  \bibnamefont{and} \bibinfo{author}{\bibfnamefont{J.~M.~M.}
  \bibnamefont{Senovilla}}, \bibinfo{journal}{Gen. Relativ. Gravit.}
  \textbf{\bibinfo{volume}{29}}, \bibinfo{pages}{91} (\bibinfo{year}{1997}),
  ISSN \bibinfo{issn}{0001-7701},
  \urlprefix\url{http://link.springer.com/10.1023/A:1010256231517}.

\bibitem[{\citenamefont{Clifton et~al.}(2013)\citenamefont{Clifton, Ellis, and
  Tavakol}}]{Clifton2013}
\bibinfo{author}{\bibfnamefont{T.}~\bibnamefont{Clifton}},
  \bibinfo{author}{\bibfnamefont{G.~F.~R.} \bibnamefont{Ellis}},
  \bibnamefont{and} \bibinfo{author}{\bibfnamefont{R.}~\bibnamefont{Tavakol}},
  \bibinfo{journal}{Class. Quantum Gravity} \textbf{\bibinfo{volume}{30}},
  \bibinfo{pages}{125009} (\bibinfo{year}{2013}), ISSN
  \bibinfo{issn}{0264-9381},
  \urlprefix\url{http://stacks.iop.org/0264-9381/30/i=12/a=125009?key=crossref.6a5451930f226fae7034e7d503b56efd}.

\bibitem[{\citenamefont{Sussman and Larena}(2014)}]{Sussman2014}
\bibinfo{author}{\bibfnamefont{R.~A.} \bibnamefont{Sussman}} \bibnamefont{and}
  \bibinfo{author}{\bibfnamefont{J.}~\bibnamefont{Larena}},
  \bibinfo{journal}{Class. Quantum Gravity} \textbf{\bibinfo{volume}{31}},
  \bibinfo{pages}{075021} (\bibinfo{year}{2014}), ISSN
  \bibinfo{issn}{0264-9381},
  \urlprefix\url{http://stacks.iop.org/0264-9381/31/i=7/a=075021?key=crossref.54b06cb39360335b92ba919adb244c35}.

\bibitem[{\citenamefont{Faraoni}(2007)}]{Faraoni2007}
\bibinfo{author}{\bibfnamefont{V.}~\bibnamefont{Faraoni}},
  \bibinfo{journal}{Phys. Rev. D} \textbf{\bibinfo{volume}{76}},
  \bibinfo{pages}{127501} (\bibinfo{year}{2007}), ISSN
  \bibinfo{issn}{1550-7998},
  \urlprefix\url{https://link.aps.org/doi/10.1103/PhysRevD.76.127501}.

\bibitem[{\citenamefont{Bertolami and
  Sequeira}(2009{\natexlab{b}})}]{Bertolami2009a}
\bibinfo{author}{\bibfnamefont{O.}~\bibnamefont{Bertolami}} \bibnamefont{and}
  \bibinfo{author}{\bibfnamefont{M.~C.} \bibnamefont{Sequeira}},
  \bibinfo{journal}{Phys. Rev. D} \textbf{\bibinfo{volume}{79}},
  \bibinfo{pages}{104010} (\bibinfo{year}{2009}{\natexlab{b}}), ISSN
  \bibinfo{issn}{1550-7998},
  \urlprefix\url{https://link.aps.org/doi/10.1103/PhysRevD.79.104010}.

\bibitem[{\citenamefont{Appleby et~al.}(2010)\citenamefont{Appleby, Battye, and
  Starobinsky}}]{Appleby2010}
\bibinfo{author}{\bibfnamefont{S.~A.} \bibnamefont{Appleby}},
  \bibinfo{author}{\bibfnamefont{R.~A.} \bibnamefont{Battye}},
  \bibnamefont{and} \bibinfo{author}{\bibfnamefont{A.~A.}
  \bibnamefont{Starobinsky}}, \bibinfo{journal}{J. Cosmol. Astropart. Phys.}
  \textbf{\bibinfo{volume}{2010}}, \bibinfo{pages}{005} (\bibinfo{year}{2010}),
  ISSN \bibinfo{issn}{1475-7516},
  \urlprefix\url{http://stacks.iop.org/1475-7516/2010/i=06/a=005?key=crossref.c9666ff75242b59a26a8953fc3796990}.

\bibitem[{\citenamefont{Motohashi et~al.}(2011)\citenamefont{Motohashi,
  Starobinsky, and Yokoyama}}]{Motohashi2011}
\bibinfo{author}{\bibfnamefont{H.}~\bibnamefont{Motohashi}},
  \bibinfo{author}{\bibfnamefont{A.~A.} \bibnamefont{Starobinsky}},
  \bibnamefont{and} \bibinfo{author}{\bibfnamefont{J.}~\bibnamefont{Yokoyama}},
  \bibinfo{journal}{J. Cosmol. Astropart. Phys.}
  \textbf{\bibinfo{volume}{2011}}, \bibinfo{pages}{006} (\bibinfo{year}{2011}),
  ISSN \bibinfo{issn}{1475-7516},
  \urlprefix\url{http://stacks.iop.org/1475-7516/2011/i=06/a=006?key=crossref.8e3ec3a865ed72385b73c496caff9055}.

\bibitem[{\citenamefont{Acquaviva et~al.}(2018)\citenamefont{Acquaviva,
  Kofroň, and Scholtz}}]{Acquaviva2018}
\bibinfo{author}{\bibfnamefont{G.}~\bibnamefont{Acquaviva}},
  \bibinfo{author}{\bibfnamefont{D.}~\bibnamefont{Kofroň}}, \bibnamefont{and}
  \bibinfo{author}{\bibfnamefont{M.}~\bibnamefont{Scholtz}},
  \bibinfo{journal}{Class. Quantum Gravity} \textbf{\bibinfo{volume}{35}},
  \bibinfo{pages}{095001} (\bibinfo{year}{2018}), ISSN
  \bibinfo{issn}{0264-9381},
  \urlprefix\url{http://stacks.iop.org/0264-9381/35/i=9/a=095001?key=crossref.2890b1ab7e702d63693fac660210d9aa}.

\end{thebibliography}

\end{document}